\documentclass{appolb}
\usepackage{graphicx}
\usepackage{amssymb}
\usepackage{amsbsy}
\usepackage{amsmath}
\usepackage{hyperref}
\usepackage{xcolor}
\newcommand{\nn}{\nonumber}
\newcommand{\tg}{\tilde{g}}

\begin{document}
\title{Pole properties of a resonance: When to subtract partial-decay widths to obtain the pole widths%
  \thanks{Presented at Excited QCD 2024, Benasque, Spain, 14-20 January,  2024}%
}
\author{J. A. Oller
\address{Departamento de F\'{\i}sica, Universidad de Murcia, E-30071 Murcia, Spain}
\\[3mm]
}
\maketitle
\begin{abstract}
  When a resonance lies near the threshold of a heavier channel, an interesting feature can occur. 
   The paradigmatic example employed here is the scalar isoscalar $f_0(980)$ resonance that couples to the lighter $\pi\pi$ and heavier $K\bar{K}$ channels. It is shown that the decay width  is given by the sum or subtraction of the partial decay widths depending on whether the pole lies in the Riemann sheet that is contiguous with the physical one  above or below the $K\bar{K}$ threshold, respectively.  Next, we  show that the usually disregarded renormalization of  bare parameters in  Flatt\'e or energy-dependent Breit-Wigner parameterizations is essential to extract physical information. The compositeness of the $f_0(980)$ by using a Flatt\'e parameterization matched to reproduce the pole properties  obtained from Roy equations and other analytic constraints is evaluated.  
\end{abstract}

\section{Introduction}
\label{sec.240220.1}


A Flatt\'e parameterization \cite{Flatte:1976xu} is typically used  for describing a resonance that lies near a heavier threshold. Let us denote by $i=1$ and 2 the light and  heavy channels, respectively. To fix ideas think of the $f_0(980)$ and the scalar isoscalar channels $\pi\pi$ and $K\bar{K}$, in this order. Then, around the $K\bar{K}$ threshold an $S$-wave amplitude 
 is written as
\begin{eqnarray}
  \label{240220.1}
t_{ij}&=&\frac{\tg_i\tg_j}{E-E_f+i\frac{\widetilde{\Gamma}_1}{2}+\frac{i}{2}\tg_2^2\sqrt{m_2 E}}~,
\end{eqnarray}
with $E$  the total energy measured with respect to the two-kaon threshold. The kinematics for the $K\bar{K}$ channel is treated nonrelativistically.  This parameterization is determined by three {\it bare} parameters: The bare coupling $\tg_2$, the bare width $\widetilde{\Gamma}_1$ and the bare resonance mass $E_f$. 
 $\widetilde{\Gamma}_1$ is related to the bare coupling $\tg_1$  by
\begin{eqnarray}
  \label{240220.2}
  \widetilde{\Gamma}_1&=&\tg_1^2 p_1~,
\end{eqnarray}
where $p_1$ is the $\pi\pi$ momentum at the resonance mass $M_R=\Re E_R$,  and $E_R$ is the resonance pole position in Eq.~(\ref{240220.1}).
This parameterization was extensively used in Ref.~\cite{Baru:2003qq} to study the compositeness of the $f_0(980)$ and $a_0(980)$ resonances \cite{Oller:1999ag,Albaladejo:2010tj}, 
 and revisited in Ref.~\cite{Wang:2022vga}. An issue in the traditional way of analyzing Flatté parameterizations was emphasized in the latter reference.  To illustrate it, we gather  in Table~\ref{tab.240220.1} the set of Flatt\'e parameterizations for the $f_0(980)$ used in \cite{Baru:2003qq}. From left to right,  we give the original reference,  binding momentum $p_2$, $M_R=\Re(p_2^2)/m_K$, the pole width $\Gamma=-2 \Im p_2^2/m_K$, and the bare parameters. It is striking that for all cases $\Gamma\ll \widetilde{\Gamma}_1$. Related to that (as shown below), let us note that  all the poles have $\Im p_2>0$.

\begin{table}
  \begin{tabular}{lllllll}
Ref. & $p_2$\,(MeV) & $M_R$\,(MeV)  & $\Gamma$\,(MeV)     & $\widetilde{\Gamma}_1$\,(MeV) & $\tg_2^2$ & $E_f$\,(MeV) \\ 
\hline
\cite{CMD-2:1999znb} & $-65+i\,97$ & $981$ & $50.8$ & $149$ & $1.51$ & $-84.3$ \\ 
\cite{Achasov:2000ym} & $-58+i\,107$ & $975$ & $50.1$ & $196$ & $2.51$ & $-151.5$ \\ 
\cite{Achasov:2001cj} & $-84+i\,17$ & $1005$ & $11.6$ & $129$ & $1.31$ & $+4.6$ \\ 
\cite{Antonelli:2002ip} & $-69+i\,100$ & $981$ & $55.6$ & $253$ & $2.84$ & $-154$ \\ 
  \end{tabular}
  \caption{{\small Set of Flatté parameterizations for the $f_0(980)$ considered in \cite{Baru:2003qq,Wang:2022vga}. See the text for more details.}\label{tab.240220.1}}
\end{table}

Another issue stressed in Ref.~\cite{Wang:2022vga} concerns the pole determination based on Roy equations and other analytical constraints in Ref.~\cite{Garcia-Martin:2011nna}
\begin{eqnarray}
  \label{240221.1}
  2m_K+E_R&=&996\pm 7-i\,25^{+10}_{-6}\,\text{MeV}~,~~g_1=0.46\pm 0.04  ~,
\end{eqnarray}
where $g_1$ is the physical coupling to $\pi\pi$ (adopted to our normalization), obtained from the residue of the partial-wave amplitude at the $f_0(980)$ resonance pole.  
Let us notice that the partial-decay width to $\pi\pi$ can be straightforwardly calculated by taking $g_1$ from Eq.~\eqref{240221.1} into  Eq.~\eqref{240220.2},
\begin{eqnarray}
  \label{240222.5}
  \Gamma_{\pi\pi}=g_1^2 p_1=100^{+20}_{-17}\,\text{MeV}~\cite{Burkert:2022bqo}~,
\end{eqnarray}
which is around a factor 2 larger than the pole width $\Gamma$ from Eq.~\eqref{240221.1}. How can it be?

\section{Interplay with Riemann sheets}
\label{sec.240221.1}

The $f_0(980)$ lies close to the two-kaon threshold near 1~GeV. Consequently, its physical imprint is largely dependent on the Riemann sheet (RS) in which it lies.  We characterize the different RS's \cite{Oller:2019rej} by the signs of the {\it imaginary} parts of the momenta collected as $(\pm,\pm)$, with the 1st(2nd) sign for $p_1(p_2)$. In this way, $(+,+)$ is the physical or 1st RS, $(-,+)$ the 2nd RS, $(-,-)$ the third one, and $(+,-)$ the fourth RS. This is because the square root in the calculation of the momentum as a function of energy has a right-hand cut. Therefore, when crossing the real energy axis in between the $\pi\pi$ and $K\bar{K}$ thresholds the  2nd RS $(-,+)$  connects smoothly  with the physical one;  when the real axis is crossed above the $K\bar{K}$ threshold the 3rd RS $(-,-)$ is the one that connects smoothly with the physical RS.  

It is apparent from Table~\ref{tab.240220.1} that all the $f_0(980)$ poles there are in the 2nd RS, because of the positive sign of $\Im p_2$. The pole in Eq.~\eqref{240221.1} from Ref.~\cite{Garcia-Martin:2011nna} is also in the 2nd RS. This has its importance because of the term $i g_2^2 p_2/2$ in the denominator in Eq.~\eqref{240220.1}. One can consider only one complex $E$ plane 
by using the common convention in numerical calculations, like in Fortran, such that the square root has a left-hand cut.  In this way, the 1st RS corresponds to $\Im E>0$, and Eq.~\eqref{240220.1} applies. For $\Im E<0$ the expressions for the different RS's are: 
  \begin{eqnarray}
\label{240222.1}
\text{2nd RS:} & t_{ij}=& \frac{\tg_i \tg_j}{E-E_f+i\frac{\widetilde{\Gamma}_1}{2}\textcolor{red}{\mathbf{-}}\frac{i}{2}\tg_2^2\sqrt{m_2 E}}~,\\
\text{3rd RS:} & t_{ij}=&
\frac{\tg_i \tg_j}{E-E_f+i\frac{\widetilde{\Gamma}_1}{2}\textcolor{red}{\mathbf{+}}\frac{i}{2}\tg_2^2\sqrt{m_2 E}}~.\nonumber
  \end{eqnarray}
  We have stressed in Eq.~\eqref{240222.1} the sign in front of $p_2=\sqrt{m_2 E}$. 
  Then, we see that  in the 3rd RS the $K\bar{K}$ and $\pi\pi$ partial-decay widths add up, whereas in the 2nd RS $p_2$ flips its sign and the $K\bar{K}$ contribution is subtracted to the $\pi\pi$ one to get the pole width. As a result, we have the following relations between the pole width $\Gamma$ and the partial-decay widths: 
\begin{eqnarray}
  \label{240222.2}
 \text{2nd RS:~} \Gamma&=\Gamma_{\pi\pi}-\Gamma_{K\bar{K}}~,\\
  \label{240222.3}
 \text{3rd RS:~} \Gamma&=\Gamma_{\pi\pi}+\Gamma_{K\bar{K}}~. 
\end{eqnarray}
The subtraction between the partial-decay widths in Eq.~\eqref{240222.2} for the pole in the 2nd RS was first unveiled in Ref.~\cite{Wang:2022vga}.
Two corollaries  follow from Eq.~\eqref{240222.2}: 1) Since $\Gamma>0$ then
$\Gamma_{\pi\pi}> \Gamma_{K\bar{K}}$. 2) As $\Gamma_{\pi\pi}=\Gamma+\Gamma_{K\bar{K}}$ then $\Gamma_{\pi\pi}>\Gamma_{K\bar{K}}$.

The result in Eq.~\eqref{240222.2} was then applied in Ref.~\cite{Burkert:2022bqo} to the pole position of the $f_0(980)$  in the 2nd RS \cite{Garcia-Martin:2011nna} given in Eq.~\eqref{240221.1}. The fact that $\Gamma_{\pi\pi}>\Gamma$, cf.  Eq.~\eqref{240222.5}, is now understood as due to the negative contribution of $\Gamma_{K\bar{K}}$ to $\Gamma$. Thus, $\Gamma_{K\bar{K}} = \Gamma_{\pi\pi}-\Gamma=50^{+26}_{-21}~\text{MeV},$ as calculated in Ref.~\cite{Burkert:2022bqo}. Another quantity of interest that was addressed in this reference is the definition of the total width $\Gamma_{\rm tot}=\Gamma_{\pi\pi}+\Gamma_{K\bar{K}}$, which does not coincide with the pole width  $\Gamma=\Gamma_{\pi\pi}-\Gamma_{K\bar{K}}$. The definition of $\Gamma_{\rm tot}$, the same independently of the sheet in which the pole lies, reflects the fact that in an event distribution the total sum of resonant events is the sum of  events for every channel separately. We report from Ref.~\cite{Burkert:2022bqo} the following results 
\begin{eqnarray}
  \label{240223.1}
  \Gamma_{\rm tot}&=&\Gamma_{\pi\pi}+\Gamma_{K\bar{K}}=151^{+44}_{-37}~\text{MeV}~,\\
 \text{BR}_{\pi\pi}&=&\frac{\Gamma_{\pi\pi}}{\Gamma_{\rm tot}}=0.67\pm 0.07~,\nn\\
 \text{BR}_{K\bar{K}}&=&\frac{\Gamma_{K\bar{K}}}{\Gamma_{\rm tot}}=0.33\pm 0.07~,\nn\\
 r_{K\bar{K}/\pi\pi}&=&\frac{\Gamma_{K\bar{K}}}{\Gamma_{\pi\pi}}=0.49\pm 0.11~.\nn
\end{eqnarray}

\section{Renormalization of bare parameters in phenomenological parameterizations}
\label{sec.240223.1}

Let us illustrate the process for the Flatt\'e parameterization in Eq.~\eqref{240220.1}, though the same line of argumentation could be applied to an energy dependent Breit-Wigner as well.
We refer to \cite{Wang:2022vga} for the more detailed and original derivation.

The  pole position $E_R$ in Eq.~\eqref{240220.1} near to the $K\bar{K}$ threshold is  \cite{Wang:2022vga}
\begin{eqnarray}
  \label{240223.2}
  E_R&=E_f-\frac{m_K}{8}g_2^4-\frac{i}{2}\widetilde{\Gamma}_{\pi\pi}+\sigma \frac{g_2^2}{2}\sqrt{m_K\left(\frac{m_K g_2^4}{16}-E_f+\frac{i}{2}\widetilde{\Gamma}_{\pi\pi}\right)}~,
  \end{eqnarray}
where $\sigma=+1(-1)$ corresponds to the pole lying the 2nd(3rd) RS. Next, it is important to calculate the behavior of the denominator of $t_{ij}(E)$ in Eq.~\eqref{240220.1} for $E\to E_R$. One has that \cite{Wang:2022vga} 
\begin{eqnarray}
  \label{240223.3}
  \beta&\equiv& \lim_{E\to E_R}\frac{E-E_R}{E-E_f+i\frac{\widetilde{\Gamma}_1}{2}+\frac{i}{2}g_2^2\sqrt{m_2 E}}\nn\\
&=&4\sqrt{|E_R|}{m_k g_2^4+16|E_R|+4\sigma g_2^2\sqrt{2m_K(|E_R|-M_R)}}~,
\end{eqnarray}
with $|E_R|=\sqrt{M_R^2+\Gamma^2/4}$. 
 Therefore, the renormalized or physical couplings in a Flatt\'e parameters are
\begin{eqnarray}
\label{240223.6}
  g_i=\beta^\frac{1}{2} \tilde{g}_i~,
\end{eqnarray}
such that the physical width to $\pi\pi$ is $\Gamma_{\pi\pi}=\beta \widetilde{\Gamma}_{\pi\pi}$. The renormalized couplings $g_i$   are the ones that must be compared with the couplings obtained by evaluating the residue of a $T$-matrix, like $g_1$ given in Eq.~\eqref{240221.1} from Ref.~\cite{Garcia-Martin:2011nna}.  It is important to stress this point because it is common in the literature to use the bare couplings and widths of a Flatt\'e parameterization as physical ones. To see the dramatic impact of $\beta$ we show in Table~\ref{tab.240223.1} the values of $\beta$, $\Gamma_{\pi\pi}=\beta\widetilde{\Gamma}_{\pi\pi}$ and $\Gamma_{K\bar{K}}=\Gamma_{\pi\pi}-\Gamma$ that correspond to the same Flatt\'e analyses as in Table~\ref{tab.240220.1}.  It is obvious from the new results in Table~\ref{tab.240223.1} that $\widetilde{\Gamma}_{\pi\pi}$ is very different to the physical one $\Gamma_{\pi\pi}$, and that $\Gamma_{\pi\pi}>\Gamma$. 
 Thus,  Eqs.~\eqref{240222.2} and \eqref{240223.6} allow to properly extract the physical couplings and widths from a Flatté parameterization.

\begin{center}
  \begin{table}
  \begin{tabular}{lllllll}
    Ref. & $p_2$\,(MeV) & $\Gamma$\,(MeV) & $\beta$ & $\Gamma_{\pi\pi}=\beta \widetilde{\Gamma}_{\pi\pi}$\,(MeV) & $\Gamma_{K\bar{K}}=\Gamma_{\pi\pi}-\Gamma$ \\
    \hline
    \cite{CMD-2:1999znb} & $-65+i\,97$ & $50.8$ & 0.40 & 59.7      & 8.9 \\
    \cite{Achasov:2000ym} & $-58+i\,107$ & $50.1$ & 0.29 & 56.5    & 6.4 \\
    \cite{Achasov:2001cj} & $-84+i\,17$ & $11.6$ & 0.27 & 67.2     & 11.6 \\
    \cite{Antonelli:2002ip} & $-69+i\,100$ &  $55.6$ & 0.43 & 55.7 &  44.1 \\
  \end{tabular}
  \caption{{\small The $\beta$ parameter and partial decay widths for the Flatt\'e parameterizations in Table~\ref{tab.240220.1}. Notice the smallness of $\beta$.}\label{tab.240223.1}}
  \end{table}
\end{center}


\subsection{Compositeness analysis}
\label{sec.240227.1}

Now, let us apply the compositeness relation from Refs.\cite{Guo:2015daa,Oller:2017alp} to the $f_0(980)$ pole from Ref.~\cite{Garcia-Martin:2011nna}
\begin{align}
  \label{240227.1}
  X&=X_1+X_2~,~
  X_1=\gamma_1^2\left|\frac{\partial G_1}{\partial s}\right|_{s_R}~,~
  X_2=\gamma_2^2\left|\frac{\partial G_2}{\partial s}\right|_{s_R}~.
\end{align}
In this equation  $X_i$ is the partial compositeness of channel $i$ and $X$ is the total compositeness. Let us recall that the a partial compositeness is the weight of this channel in the composition of the state, and the total compositeness is the total weight of the meson-meson components. Regarding the different ingredients in Eq.~\eqref{240227.1}: 1) $s$ is the Mandelstam variable $s=P^2$, with $P$ the total four-momentum, and $s_R=(2m_K+E_R)^2$. 2) The couplings $\gamma_i$ are just proportional to $g_i$, such that $\gamma_i= g_i \sqrt{8\pi \Re s_R}$. 3) The functions $G_i(s)$ are the relativistic unitarity loop functions
\begin{align}
  \label{240227.2}
  G_i&
  =-\frac{1}{16\pi^2}\ln\frac{\sigma(s)-1}{\sigma(s)+1}~,~\sigma(s)=\sqrt{1-\frac{4m_i^2}{s}}~.
\end{align}
To establish a Flatt\'e parameterization requires three parameters $(\tg_2,\widetilde{\Gamma}_{\pi\pi},E_f)$. From the pole position \cite{Garcia-Martin:2011nna} in Eq.~\eqref{240221.1} we  can fix two parameters, but one more is still necessary. Then, as in Ref.~\cite{Wang:2022vga}, we take $X$ as the third input, and  calculate as a function of it the physical quantities. 
It turns out that only for $X>0.6$ the value of $g_1$ that results is compatible with  Eq.~\eqref{240221.1}. In more detail, we have $(X,g_1)=(1,0.47)$, $(0.8,0.45)$, and $(0.6,0.42)$. 
 In the same order, $X_2=0.96$, 0.76 and 0.57, respectively, with $X_1\lesssim 0.04\ll X_2$ for all cases. Therefore, the application of the compositeness relation of Eq.~\ref{240227.1} to a Flatt\'e parameterization required to reproduce the pole properties of the $f_0(980)$ from Ref.~\cite{Garcia-Martin:2011nna}, drives to the conclusion that this pole is mainly of composite nature, with its composition dominated by the $K\bar{K}$ contribution.

 In summary, we have shown that the pole width for a pole lying in the 2nd RS is given by the subtraction of the partial-decay widths. We have also discussed the renormalization process of the bare parameters in  Flatt\'e or energy-dependent Breit-Wigner parameterizations. 
 For clarification and to avoid confusion, phenomenological analyses should provide the renormalized  parameters, the physically meaningful ones,  which can be worked out straightforwardly from the bare ones directly employed in these parameterizations. 

 \vskip 5pt
 {\bf Acknowledgements.} I would like to thank partial financial support to the Grant PID2022-136510NB-C32 funded by\\ MCIN/AEI/10.13039/501100011033/ and FEDER, UE.


\bibliographystyle{unsrt}
\bibliography{references}

\end{document}